\documentclass[12pt]{article}

\pdfoutput=1

\usepackage{graphicx}
\usepackage{dcolumn}
\usepackage{bm}
\usepackage{hyperref}
\usepackage{color}
\usepackage{amsmath}
\usepackage{amssymb}
\usepackage{amsfonts}
\usepackage{multirow}
\usepackage{cite}

\usepackage[top=30truemm,bottom=30truemm,left=25truemm,right=25truemm]{geometry}

\begin{document}


\def\a{\alpha}
\def\b{\beta}
\def\c{\varepsilon}
\def\d{\delta}
\def\e{\epsilon}
\def\f{\phi}
\def\g{\gamma}
\def\h{\theta}
\def\k{\kappa}
\def\l{\lambda}
\def\m{\mu}
\def\n{\nu}
\def\p{\psi}
\def\q{\partial}
\def\r{\rho}
\def\s{\sigma}
\def\t{\tau}
\def\u{\upsilon}
\def\v{\varphi}
\def\w{\omega}
\def\x{\xi}
\def\y{\eta}
\def\z{\zeta}
\def\D{\Delta}
\def\G{\Gamma}
\def\H{\Theta}
\def\L{\Lambda}
\def\F{\Phi}
\def\P{\Psi}
\def\S{\Sigma}

\def\o{\over}
\def\beq{\begin{align}}
\def\eeq{\end{align}}
\newcommand{\gsim}{ \mathop{}_{\textstyle \sim}^{\textstyle >} }
\newcommand{\lsim}{ \mathop{}_{\textstyle \sim}^{\textstyle <} }
\newcommand{\vev}[1]{ \left\langle {#1} \right\rangle }
\newcommand{\bra}[1]{ \langle {#1} | }
\newcommand{\ket}[1]{ | {#1} \rangle }
\newcommand{\EV}{ {\rm eV} }
\newcommand{\KEV}{ {\rm keV} }
\newcommand{\MEV}{ {\rm MeV} }
\newcommand{\GEV}{ {\rm GeV} }
\newcommand{\TEV}{ {\rm TeV} }
\newcommand{\1}{\mbox{1}\hspace{-0.25em}\mbox{l}}
\newcommand{\headline}[1]{\noindent{\bf #1}}
\def\diag{\mathop{\rm diag}\nolimits}
\def\Spin{\mathop{\rm Spin}}
\def\SO{\mathop{\rm SO}}
\def\O{\mathop{\rm O}}
\def\SU{\mathop{\rm SU}}
\def\U{\mathop{\rm U}}
\def\Sp{\mathop{\rm Sp}}
\def\SL{\mathop{\rm SL}}
\def\tr{\mathop{\rm tr}}
\def\mpl{M_{\rm Pl}}

\def\IJMP{Int.~J.~Mod.~Phys. }
\def\MPL{Mod.~Phys.~Lett. }
\def\NP{Nucl.~Phys. }
\def\PL{Phys.~Lett. }
\def\PR{Phys.~Rev. }
\def\PRL{Phys.~Rev.~Lett. }
\def\PTP{Prog.~Theor.~Phys. }
\def\ZP{Z.~Phys. }

\def\dd{\mathrm{d}}
\def\ff{\mathrm{f}}
\def\BH{{\rm BH}}
\def\inf{{\rm inf}}
\def\ev{{\rm evap}}
\def\eq{{\rm eq}}
\def\SM{{\rm sm}}
\def\Mpl{M_{\rm Pl}}
\def\GeV{{\rm GeV}}

\def\Msusy{m_{\rm stop}}
\newcommand{\Red}[1]{\textcolor{red}{#1}}
\newcommand{\TL}[1]{\textcolor{blue}{\bf TL: #1}}

\baselineskip 0.7cm
\begin{titlepage}

\vskip 1.35cm
\begin{center}

{\bf\large Impact of an extra gauge interaction \\ on naturalness of supersymmetry}

\vskip 1.2cm
Marcin Badziak$^{1}$ and Keisuke Harigaya$^{2,3}$
\vskip 0.4cm
$^1${\it Institute of Theoretical Physics, Faculty of Physics, University of Warsaw, ul.~Pasteura 5, PL--02--093 Warsaw, Poland}\\
$^2${\it Department of Physics, University of California, Berkeley, California 94720, USA}\\
$^3${\it Theoretical Physics Group,  Lawrence Berkeley National Laboratory, Berkeley, California 94720, USA}
\vskip 1.5cm

\abstract{
It is pointed out that in supersymmetric models with a new gauge symmetry under which the Higgs is charged, the fine-tuning of
the electroweak symmetry breaking is relaxed due to suppression of the top Yukawa coupling at higher scales by a new large gauge coupling.
We calculate the fine-tuning in an explicit model and find that the lower bounds on stops and gluino masses from the naturalness criterion are
increased by several hundred GeV in comparison to the Minimal Supersymmetric Standard Model (MSSM). 
The fine-tuning is improved by one to two orders of
magnitude as compared to the MSSM, as a consequence of both the suppression of the top Yukawa coupling and
the additional
tree-level contribution to the Higgs mass allowing for much lighter stops.
}

\end{center}
\end{titlepage}

\setcounter{page}{2}

\section{Introduction}

Supersymmetry (SUSY) provides one of the best solutions to the hierarchy problem of the Standard Model (SM)~\cite{MaianiLecture,Veltman:1980mj,Witten:1981nf,Kaul:1981wp}. However, the minimal version of
the realization of this idea, the Minimal Supersymmetric Standard Model (MSSM), has been put under a strong pressure by the LHC results. The primary
problem of the MSSM is the fact that explaining the 125 GeV Higgs mass~\cite{Higgsmass_exp} requires multi-TeV stops which implies large fine-tuning
of the
electroweak (EW) scale, the so-called
little hierarchy problem. This has motivated a great amount of work on various extensions of the MSSM in which new positive contributions to the Higgs
mass are
present which allow for the 125 GeV Higgs mass without heavy stops. The most notable examples of such models are Next-To-Minimal Supersymmetric
Standard Model
(NMSSM)~\cite{reviewEllwanger,Ellwanger:2011aa,Hall:2011aa,Cao:2012fz,Jeong:2012ma,Agashe:2012zq,NaturalNMSSM_King,Gherghetta:2012gb,Badziak:2013bda}
and
models with non-decoupling $D$-terms of some new gauge interactions under which the Higgs is
charged~\cite{Batra:2003nj,Endo:2011gy,Cheung:2012zq,Huo:2012tw,DAgnolo:2012vzj,Craig:2012bs,Bertuzzo:2014sma,Capdevilla:2015qwa}.
However, even if the observed Higgs mass is obtained with light stops the fine-tuning is still present due to  constraints
on supersymmetric particles, especially on stops and gluino, from direct LHC searches~\cite{Buckley:2016kvr,Buckley:2016tbs}. 

The fine-tuning is particularly large for models with high mediation scale
of SUSY breaking, thus disfavouring many simple mechanisms of SUSY breaking such as gravity mediation. This motivated construction of new models with
extremely low  scale of SUSY breaking, see~e.g.~\cite{Dimopoulos:2014aua,Garcia:2015sfa}, models with Dirac
gauginos~\cite{Fox:2002bu,Arvanitaki:2013yja} which do not have logarithmicly-enhanced contributions to the Higgs mass parameter, and SUSY
models~\cite{Craig:2013fga,Katz:2016wtw,Badziak:2017syq,Badziak:2017kjk,Badziak:2017wxn} which suppress the fine-tuning and enhance the tree-level
Higgs mass by invoking the Twin Higgs mechanism~\cite{Chacko:2005pe,Falkowski:2006qq,Chang:2006ra}.

In the present paper, we point out that in a class of SUSY models with a new gauge symmetry under which the Higgs is charged, 
the top Yukawa coupling becomes small at high energy scales by the renormalization from the new gauge interaction. As a result, the
logarithmically-enhanced quantum correction to the Higgs mass parameter is suppressed, making models
with high mediation scale of SUSY breaking not less motivated from the naturalness perspective than models with a low mediation scale.
This mechanism is generic as long as the new gauge coupling is large, which requires the new interaction to be asymptotically free to stay in the perturbative regime up to high scales.
As a demonstraton of this idea, we quantify  in detail naturalness of the EW symmetry breaking (EWSB) in a specific model with $SU(2)_X \times
SU(2)_W$ gauge
symmetry which is broken to the SM
$SU(2)_L$ gauge group at a scale of around 10~TeV. We find that the fine-tuning may be relaxed by a factor of few as compared to the MSSM and
particularly large improvement is found in the inverted sfermion mass hierarchy scenario in which the first two generations of sfermions are much
heavier than the third one. Moreover, after the Higgs mass constraint is taken into account the improvement in the tuning, as compared to the MSSM,
is one to two orders of magnitude.

The article is organized as follows. In section~\ref{sec:2} we discuss how the suppression of the top Yukawa coupling is obtained and introduce a
model in which this scenario can be realized. In section~\ref{sec:3} we quantify the fine-tuning of the model and compare it to the MSSM. We reserve
section~\ref{sec:concl} for the summary and concluding remarks.

\section{Supersymmetry with an extra gauge symmetry}
\label{sec:2}

In this section we describe how in a model with an extra gauge symmetry the quantum correction to the Higgs mass parameter from stops and
gluino is suppressed, and introduce an explicit example.

\subsection{Suppression of the top Yukawa by an extra gauge interaction}
In the MSSM, the soft mass of the Higgs $m_{H_u}^2$ receives a large quantum correction from those of stops $m_{Q_3}^2$, $m_{U_3}^2$ because of the
large top Yukawa coupling $y_t$,
\begin{align}
\frac{{\rm d}}{{\rm d ln}\mu_R} m_{H_u}^2 =&  \frac{6 y_t^2}{16\pi^2}\left(m_{Q_3}^2 + m_{U_3}^2 \right)  + \cdots.
\end{align}
The effect is especially significant for a large mediation scale of SUSY breaking because of the large logarithmic enhancement, which leads to an
excessive amount of the fine-tuning to obtain the EWSB scale.

We point out that this problem is relaxed in extensions of the MSSM such that the Higgs is charged under extra gauge symmetry,
which have been extensively discussed becuase they alllow  for the 125 GeV Higgs mass with relatively small stop
masses~\cite{Batra:2003nj,Endo:2011gy,Cheung:2012zq,Huo:2012tw,DAgnolo:2012vzj,Craig:2012bs,Bertuzzo:2014sma,Capdevilla:2015qwa}.
The running of the top Yukawa coupling is given by
\begin{align}
\frac{{\rm d}}{{\rm dln} \mu_R} y_t =  \left( \gamma_{H_u} + \gamma_{Q_3} + \gamma_{\bar{U}_3} \right) y_t,
\end{align}
where $\gamma_i$ is the anomalous dimension of the field $i$. 
At one-loop level gauge interactions gives a negative contribution to anomalous dimensions, suppressing the top Yukawa coupling at high energy
scales. As a result the quantum correction to $m_{H_u}^2$ is also suppressed.

We are interested in a model which remains perturbative up to a high energy scale and does not require a UV completion below the energy scale of
gravity.
In order to make the suppression effective by a large gauge coupling, the extra gauge interaction should be asymptotically free. We introduce an
example of such a model in the next subsection.

\subsection{A model with an extra $SU(2)$ symmetry}

We borrow the setup described in Ref.~\cite{Batra:2003nj}.
The gauge symmetry of the theory is $SU(2)_X \times SU(2)_W \times U(1)_Y \times SU(3)_c$. The chiral multiplets are shown in Table~\ref{tab:matter}.
Among quark and leptons, the first and the second generation doublets are charged under $SU(2)_W$, while the third generation doublets are charged
under $SU(2)_X$.
The Higgs multiplets $H_u$ and $H_d$ are charged under $SU(2)_X$, and their vacuum expectation values (VEVs) give masses to the third generation
quark/lepton.
The masses of the first and the second generations are given by the VEVs of $\phi_u$ and $\phi_d$. We are interested in the case where the gauge
coupling constant of $SU(2)_X$, $g_X$, is much larger than that of $SU(2)_W$, $g_W$ (``$W$" stands for ``weak", and ``$X$" does for ``extra"). The
gauge coupling $g_X$ is asymptotically free.

\begin{table}[htp]
\caption{The matter content of the model.}
\begin{center}
\begin{tabular}{|c|c|c|c|c|}
\hline
                          &$SU(2)_X$&$SU(2)_W$ & $U(1)_Y$ &$SU(3)_c$   \\ \hline
$H_u$          & ${\bf 2}$    &                  &  $1/2$     &               \\
$H_d$          & ${\bf 2}$    &                  &  $-1/2$     &               \\
$S $          & ${\bf 2}$   &   ${\bf 2}$  &                  &                   \\
$\phi_u$              &                  & ${\bf 2}$   &    $1/2$     &               \\
$\phi_{d}$     &                  &${\bf 2}$   &   $-1/2$    &                   \\
$Q_{3}$          &       ${\bf 2}$   &               &   $1/6$     &    ${\bf 3}$        \\
$Q_{1,2}$          &                 &  ${\bf 2}$   &   $1/6$     &    ${\bf 3}$        \\
$\bar{u}_{1,2,3}$     &                  &                     & $-2/3$    &        ${\bf \bar{3}}$                  \\
$\bar{e}_{1,2,3}$ &                  &                &   $1$        &                              \\
$\bar{d}_{1,2,3}$ &                  &                 & $1/3$      &     ${\bf \bar{3}}$                    \\
$L_{3}$          &    ${\bf 2}$  &                  &    $-1/2$    &                \\ 
$L_{1,2}$          &                  &  ${\bf 2}$  &    $-1/2$    &                \\ \hline
\end{tabular}
\end{center}
\label{tab:matter}
\end{table}%

The VEV of the bi-fundamental field $S$ breaks $SU(2)_X\times SU(2)_W$ down to the diagonal subgroup $SU(2)_L$, and the low energy theory is given by the MSSM.
In order to break the symmetry we consider the superpotential 
\begin{align}
W = \kappa \Xi (S_{12} S_{21} - M^2),
\end{align}
where $\Xi$ is a singlet chiral multiplet and $\kappa$, $M$ are constants, and the soft mass
\begin{align}
V_{\rm soft} = m_{S}^2|S|^2.
\end{align}
The VEV of $S$ is given by
\begin{align}
\vev{S} =
\begin{pmatrix}
0 & v_{S} \\
v_{S} & 0
\end{pmatrix},~~
v_{S} = \sqrt{M^2 - m^2 / \kappa^2}.
\end{align}
The gauge coupling of $SU(2)_L$ is 
\begin{align}
\frac{1}{g^2} = \frac{1}{g_X^2} + \frac{1}{g_W^2} \simeq \frac{1}{g_W^2},
\end{align}
and the mass of the gauge boson of the broken symmetry is given by
\begin{align}
m_X^2 = (g_X^2 + g_W^2) v_{S}^2.
\end{align}
The experimental lower bound of $m_X$ is $m_X\gtrsim g_X \times4.1$~TeV~\cite{Badziak:2017kjk}.
After integrating out the field $S$, the effective quartic coupling arising from the  $SU(2)_X \times SU(2)_W$ gauge symmetry is given by
\begin{align}
V_{\rm eff} =
\left (\frac{g^2}{8} + \delta \lambda \right) (|H_u|^2 - |H_d|^2)^2 , \\
\delta \lambda = \frac{(g_X^2 - g^2) m_S^2}{ 4 ( m_X^2 + 2 m_S^2)}
\simeq \frac{g_X^2 }{ 8} (1-\epsilon^2)\ ,~~
\epsilon^2 \equiv
\frac{m_X^2}{m_X^2 + 2 m_S^2}\ ,
\end{align}

In order to obtain a quartic coupling larger than that of the MSSM, the soft mass $m_S^2$ must be non-zero or, equivalently, $\epsilon^2<1$. The
one-loop threshold correction around the symmetry breaking scale generates a soft mass of the Higgs, which in the limit $g_X^2 \gg g_W^2 $ is given by
\begin{align}
\label{eq:thr}
\delta m_{H_u}^2 \simeq 3 \frac{g_X^2}{64\pi^2} m_X^2 {\rm ln} \left( \epsilon^{-2} \right).
\end{align}
This threshold correction shifts also the soft masses of all other $SU(2)_X$-charged fields.

The Higgs mass in this model is given by
\begin{equation}
 m_h^2= (M_Z^2 + 4\delta \lambda v^2) \cos^2\left(2\beta\right) + \left(\delta m_h^2 \right)_{\rm loop} \,,
\end{equation}
where $v\approx 174$~GeV and we assumed the decoupling limit in which the mass of the heavy Higgs doublets $m_H \gg m_h$.  $\left(\delta m_h^2
\right)_{\rm loop}$ parameterizes the loop correction which, similarly as in the MSSM, is dominated by loops involving stops. In the MSSM, for the
stop masses of 1 (2) TeV without stop mixing the Higgs mass is only about 110 (115) GeV even for large $\tan\beta$ \cite{susyHD}. In the present model
the 125 GeV Higgs mass can be easily obtained for any stop mass by appropriate choice of $g_X$, $\epsilon^2$ and $\tan\beta$.   
We consider a case with a large gauge coupling of the new interaction, $g_X\gtrsim2$, so the 125 GeV Higgs mass requires values of
$\epsilon^2$
close to unity e.g. for the stop masses of 1 (2) TeV
without stop
mixing and $g_X=2.5$ the 125 GeV Higgs mass requires $\epsilon^2\approx0.96\ (0.98)$ for large $\tan\beta$.

\begin{figure}[t]
\centering
\includegraphics[clip,width=.48\textwidth]{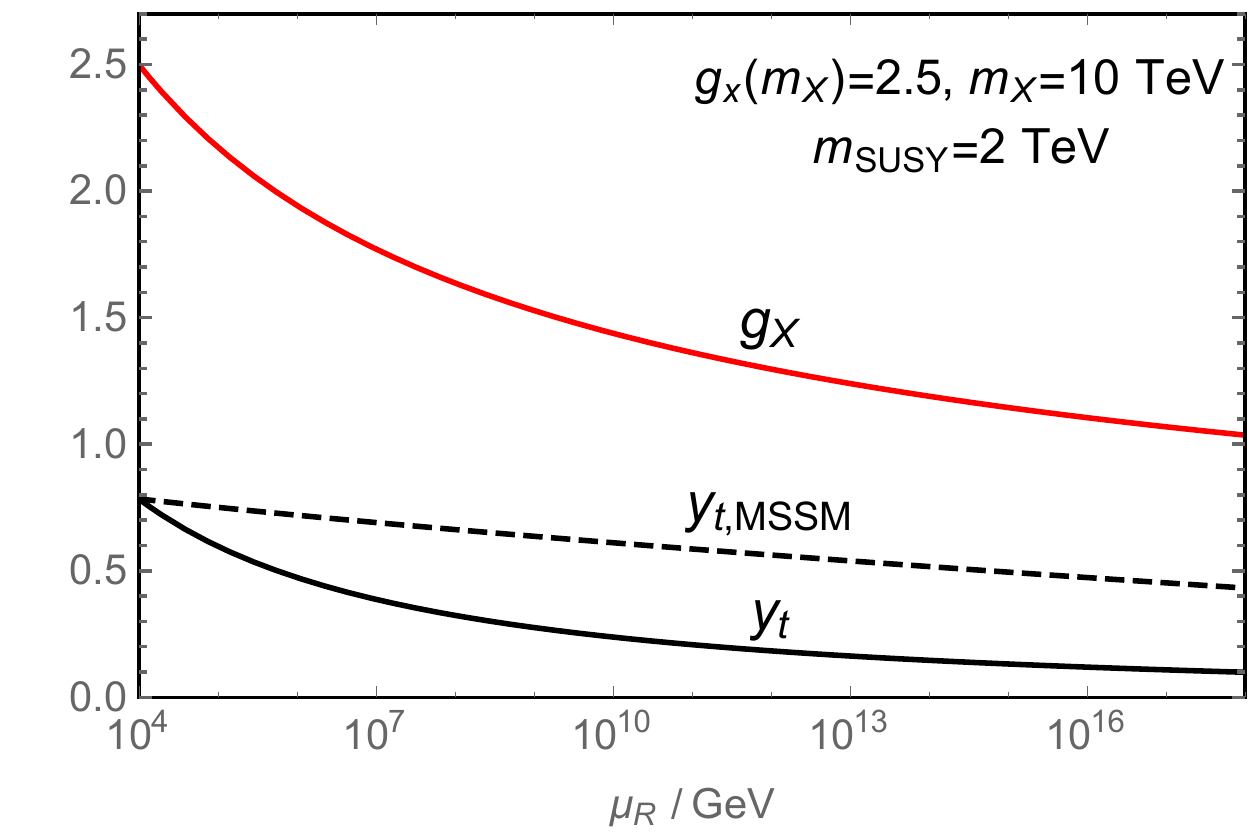}
\caption{
RG running of the gauge coupling $g_X$ (red) and the top Yukawa coupling $y_t$ for $m_X=10$ TeV, $m_{\rm SUSY}=2$ TeV, $g_X(m_X)=2.5$.
}
\label{fig:run}
\end{figure}

The renormalization group (RG) running of the top Yukawa coupling is given by
\begin{align}
\frac{{\rm d}}{{\rm dln} \mu_R} y_t = \frac{y_t}{16\pi^2} \left( -3 g_X^2 - \frac{16}{3}g_3^2 - \frac{13}{15}g_1^2 + 6 y_t^2    \right).
\end{align}
In fig.~\ref{fig:run} we show the running of the gauge coupling $g_X$ and the top Yukawa coupling $y_t$. For comparison we also show the running of
the top Yukawa coupling in the MSSM. We investigate the impact of the suppressed top Yukawa coupling to the naturalness of the EWSB scale in the next
section.

\section{Naturalness of electroweak symmetry breaking}
\label{sec:3}

In this section we evaluate the naturalness of the EWSB scale of the model explained in the previous section. We quantify the fine-tuning using the
following measure 
\begin{equation}
\Delta \equiv {\rm max}\{\Delta_{x_i}\} \,, \qquad \Delta_{x_i} \equiv  |\frac{\partial{\rm ln} v^2}{\partial{\rm ln} x_i(\Lambda)}| \,,
\end{equation}
where $x_i$ is a mass-squared parameter (e.g.~$M_3^2$, $m_{Q_3}^2$, $\mu^2$) at the mediation scale of SUSY breaking $\Lambda$. $\mu^2$ is a
supersymmetric parameter and contributes to fine-tuning at tree level:
\begin{equation}
 \Delta_{\mu^2} \approx  \frac{2\mu^2}{m_h^2} \,.
\end{equation}
The soft SUSY breaking parameters contribute to fine-tuning via loop corrections to $m_{H_u}^2$. In order to compute these corrections we solve the
RG equations (RGEs) between the SUSY scale, set to 2~TeV, and $\Lambda$.
We have set  $\tan\beta=10$ and neglected all the Yukawa couplings other than the top Yukawa coupling.
The RGEs are solved basically at the one-loop level except for the followings:
Two-loop corrections to soft masses from the first two generations of sfermions are included as we will also discuss the case where they are heavy.
As $g_X$ is large, the RGEs of the gauge couplings are solved at the two-loop level. Two-loop corrections from the soft masses of $SU(2)_X$-charged chiral multiplets are also included.

Between $\Lambda$ and $m_X$, the RGEs are solved in the unbroken $SU(2)_X$ phase.
The size of the two-loop correction to $m_{H_u}^2$ from $m_S^2$, which for $\epsilon^2=0.95$ and $m_X \approx 10$ TeV is about $\left(1.5\
\TEV\right)^2$ at the $m_X$ scale,
depends on the RG running of $m_S^2$. In our analysis we assume that $m_S^2$ is constant which is approximately the case if the $SU(2)_X$ gaugino mass and
the soft mass of $\Xi$ are suppressed. In this approximation $m_S^2$ corrects $m_{H_u}^2$ by about $-(300\GEV)^2$ for $g_X=2.5$ so has subdominant
impact on the
fine-tuning. 

At a scale $m_X$ we perform matching by including the threshold correction~\eqref{eq:thr} to soft masses for all $SU(2)_X$-charged fields. The
exact size of this correction depends on the parameters determining the Higgs mass. In the numerical calculations we set for concreteness $g_X=2.5$,
$\epsilon^2\approx0.95$ and $\tan\beta=10$ which can explain the 125 GeV Higgs mass for sub-TeV stop masses even without stop mixing. 
For this choice of parameters, and  after saturating the experimental lower limit
on $m_X$ shown in the previous section, the threshold correction is about $(400\GEV)^2$ and does not affect
fine-tuning much. For larger stop masses
and/or non-negligible stop mixing the threshold correction can be even smaller due to $\epsilon^2$ being closer to unity.
Between $m_X$ and the SUSY scale we solve the MSSM RGEs.

In order to understand how the new interaction impacts the naturalness of EWSB it is instructive to express the IR value of $m_{H_u}^2$ in terms
of the most relevant UV soft SUSY breaking parameters. For the MSSM RGEs, which is a useful reference point, this relation is given by
\begin{align}
 \left(m_{H_u}^2\right)_{\rm IR}^{\rm MSSM} &\approx 0.68 m_{H_u}^2 - 0.32 m_{Q_3}^2 - 0.25 m_{U_3}^2  + 0.005 m_0(1,2)^2 \nonumber \\ 
 &-1.37 M_3^2 + 0.21
M_2^2
-0.13 M_2 M_3 \,,
\label{eq:coeff_mHuMSSM}
\end{align}
where all soft terms  on the r.h.s.~are defined at the UV scale $\Lambda=10^{16}$~GeV.
$m_0(1,2)^2$ is the soft masses of the first two generation sfermions.
The above formula clearly demonstrates the well-known fact
that fine-tuning in the MSSM is dominated by stops and gluino which all
give negative contribution to $m_{H_u}^2$ via the renormalization by the large top Yukawa coupling.

In the presence of a large new gauge coupling the top Yukawa coupling is driven to smaller values at high energy scales. In the model considered in
this
paper with $g_X=2.5$ the relation
between the IR value of $m_{H_u}^2$ and the UV parameters reads:
\begin{align}
 \left(m_{H_u}^2\right)_{\rm IR} & \approx  0.82 m_{H_u}^2 - 0.22 m_{Q_3}^2 - 0.08 m_{U_3}^2
 + 0.005 m_0(1,2)^2  \nonumber \\ 
 & -1.1 M_3^2 + 8.9 M_X^2
-1.39 M_X M_3 \,,
\label{eq:coeff_mHu}
\end{align}
where $M_X$ is the $SU(2)_X$ gaugino mass. Note that the coefficient in front of $m_{U_3}^2$ is reduced by a factor of three as compared to the MSSM case, as a consequence of suppressed top
Yukawa coupling at high energy scales.  The coefficients in front of $m_{Q_3}^2$ and $M_3^2$ are also reduced but not as much. This is due to compensating
effects induced by large $g_X$. Namely,
$m_{Q_3}^2$ gives additional negative two-loop correction to $m_{H_u}^2$ by the $SU(2)_X$ gauge interaction,
\begin{align}
\frac{{\rm d}}{{\rm d ln}\mu_R} m_{H_u}^2 = \frac{3 g_X^4}{256\pi^4} \times 3 m_{Q_3}^2 + \cdots,
\end{align}
while $M_3$ gives a negative contribution to $m_{H_u}^2$ via the Higgs-stop trilinear coupling $A_t$,
\begin{align}
\frac{{\rm d}}{{\rm d ln}\mu_R} m_{H_u}^2 =&  \frac{6 y_t^2}{16\pi^2} A_t^2 + \cdots, \nonumber \\
\frac{{\rm d}}{{\rm d ln}\mu_R} A_t =& \frac{A_t}{16\pi^2}\left( 18 y_t^2 - \frac{16}{3}g_3^2 - 3 g_X^2 \right) + \frac{g_3^2}{16\pi^2} \frac{32}{3}
M_3 + \cdots.
\end{align}
The effect of $g_X$ dominates one-loop RG running of the top trilinear coupling $A_t$ so $A_t=0$ is no longer an attractive solution (as it is the
case in the MSSM~\cite{Carena:1993bs}).
This amplifies $|A_t|$ which is renormalized by the gluino mass and substantially feeds into the $\beta$-function of $m_{H_u}^2$ via the top Yukawa
coupling (despite the latter being suppressed). As a consequence of these compensating effects of $y_t$ and $g_X$ the coefficients in front of
$m_{Q_3}^2$ and $M_3^2$ are stable as a function of $g_X$ and vary by only few percent for $g_X$ in the range between 2 an 3.  We 
set $g_X=2.5$ in the rest of our numerical analysis of the present paper. We also see from eq.~\eqref{eq:coeff_mHu} a large dependence of $m_{H_u}^2$
on the $SU(2)_X$ gaugino mass $M_X$ so we set it to zero throughout the analysis.

Fine-tuning in various supersymmetric extensions of the SM has been usually computed assuming some correlations between the UV parameters,
see~e.g.~\cite{Chankowski:1997zh,Feng:1999mn,Abe:2007kf,Horton:2009ed,Ross:2011xv,Yanagida:2013ah,Kaminska:2013mya,Harigaya:2015iva,Harigaya:2015jba,Ross:2017kjc}. In the present paper we remain
agnostic about any such correlations and take purely phenomenological point of view
and calculate the fine-tuning measure as a function of physical masses of sparticles. 
In this approach collider constraints on sparticle masses may be translated to lower bounds on $\Delta$. Such procedure was recently adopted in
Ref.~\cite{Buckley:2016tbs} using the MSSM RGEs. However, in the work of Ref.~\cite{Buckley:2016tbs} $m_{H_u}^2$ was not included in the set of
parameters $\{x_i\}$ for which the fine-tuning is calculated. We think that $\Delta_{m_{H_u}^2}$ should be included in the definition of $\Delta$ and
do it
accordingly. In fact, $\Delta_{m_{H_u}^2}$ often dominates the total amount of fine-tuning unless there is some large positive contribution to
$\left(m_{H_u}^2\right)_{\rm IR}$ that approximately cancels that from gluino and stops which allows for a small value of
$m_{H_u}^2$ at the mediation scale.

\begin{figure}[t]
\centering
\includegraphics[clip,width=.48\textwidth]{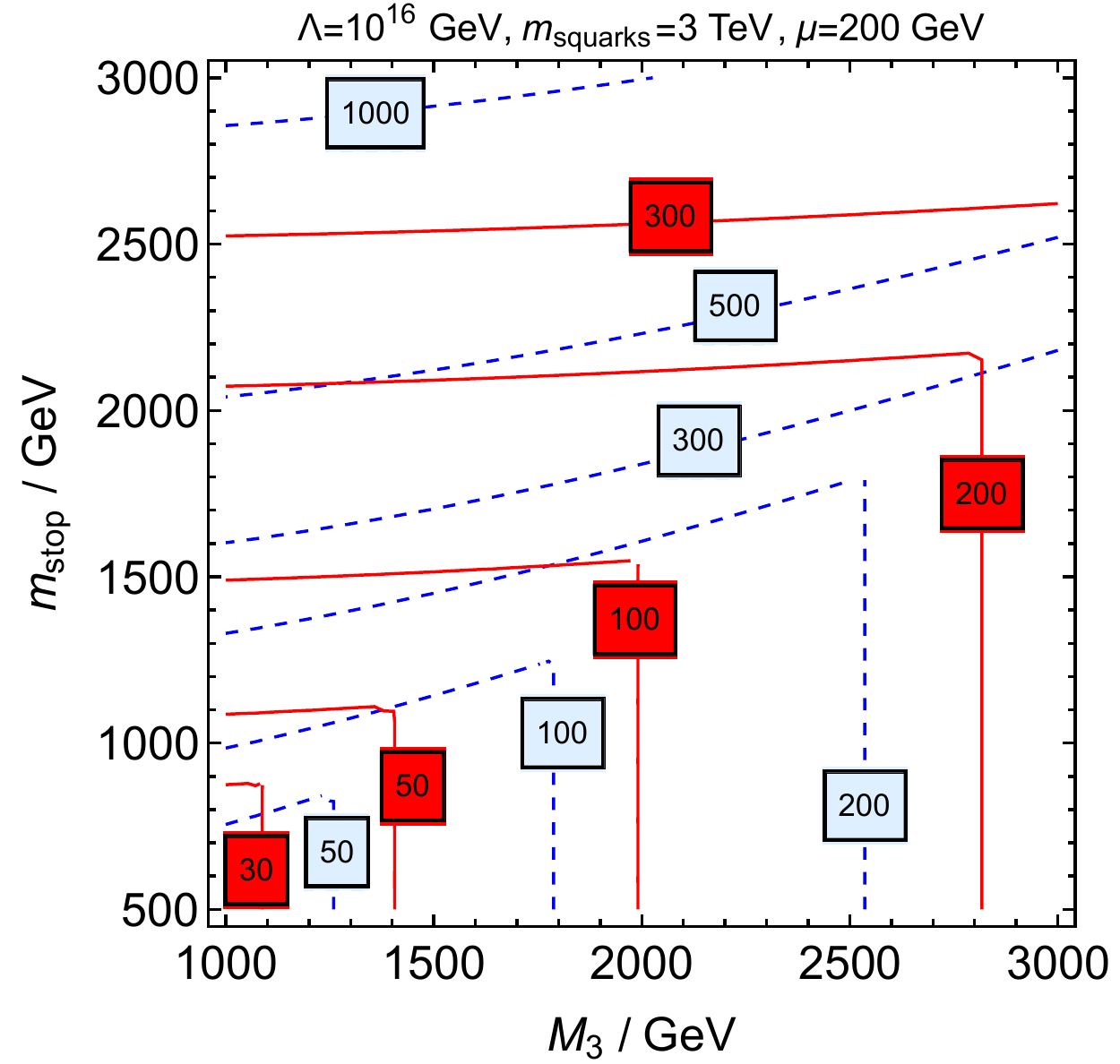}
\includegraphics[clip,width=.48\textwidth]{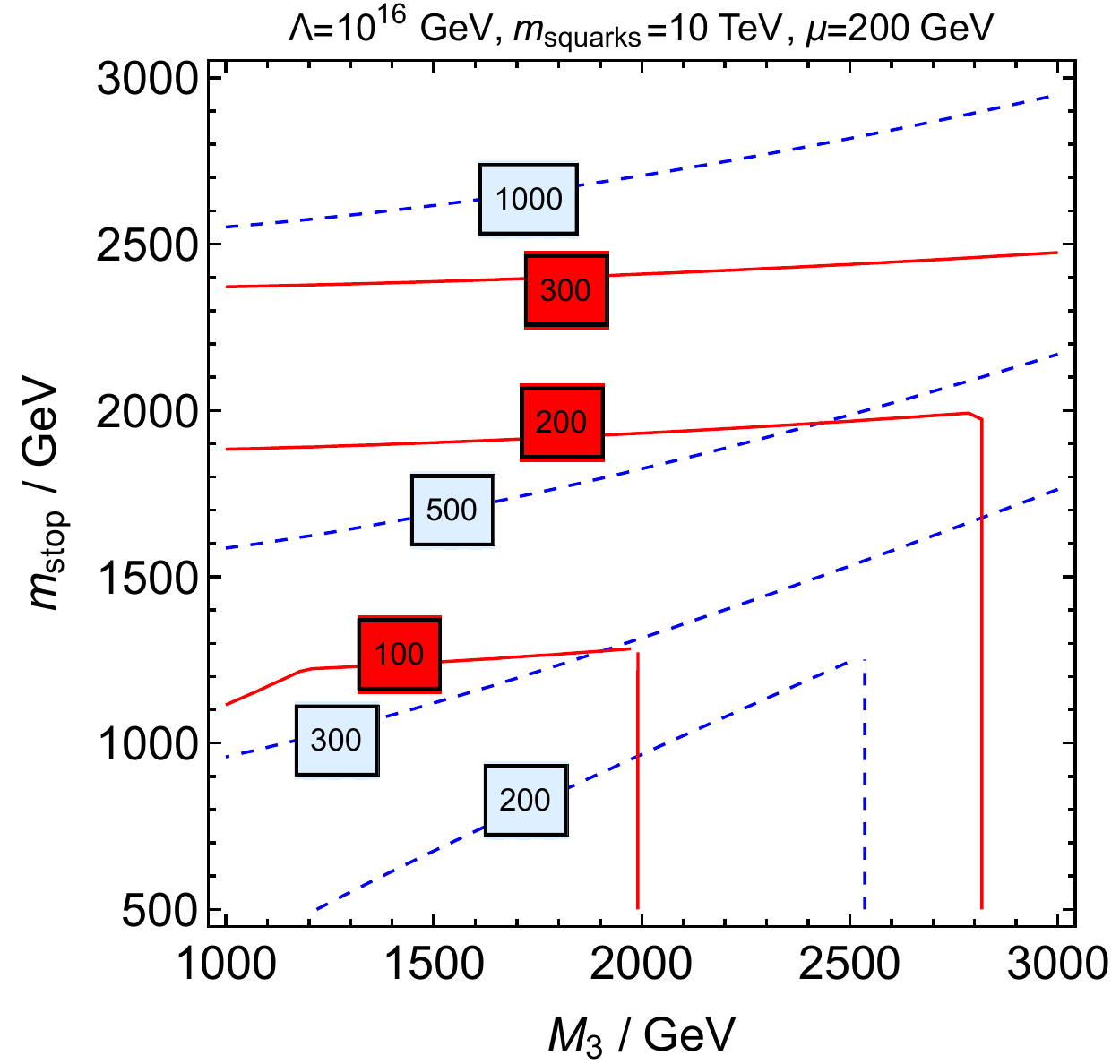}
\caption{
Contours of the fine-tuning measure $\Delta$ in the plane of IR values of the soft gluino mass $M_3$ and the mass of degenerate stops $m_{\rm
stop}$. The solid red contours corresponds to $g_X=2.5$, $\epsilon^2=0.95$ and $\tan\beta=10$. Values of $\Delta$ in MSSM are depicted by the
blue dashed contours for comparison. In the left (right) panel the sfermions of the first two generations are set to 3 (10) TeV. The remaining
parameters are set to values which do not lead to substantial corrections to the EWSB scale: $\mu=M_1=M_2=200$~GeV and the soft $SU(2)_X$ gaugino mass
$M_X=0$. 
}
\label{fig:ft_M3mstop}
\end{figure}

In the left panel of fig.~\ref{fig:ft_M3mstop} we present the fine-tuning as a function of the IR soft gluino mass and the mass of stops for
$\mu=M_1=M_2=200$~GeV. This corresponds to a generic case in which the negative corrections from stops and gluino are cancelled by a non-negligible UV
value of $m_{H_u}^2$ and $\Delta_{m_{H_u}^2}$ tends to dominate the fine-tuning. We see that for a given value of $\Delta$, gluino and stop masses
can be heavier than in
the MSSM by several hundred GeV. The tuning better than 1~\% can be obtained for a gluino
mass up to 2 TeV and stop masses up to 1.5 TeV, as compared to 1.8 and 1.2 TeV in the MSSM, respectively.  The biggest improvement in tuning is
in the region $m_{\rm stop} \gg M_3$ where corrections from stops dominate tuning and $\Delta$ is almost a factor three smaller than in the MSSM.

We also note that the improvement is particularly significant in a well-motivated scenario in which the first two generation of sfermions are much
heavier than the third one.  This scenario was suggested as a way to ease SUSY flavor
problems~\cite{Dine:1993np,Pomarol:1995xc,Dudas:1995eq,Barbieri:1995uv,Dudas:1996fe,Cohen:1996vb}. However, it
was pointed out that two-loop RGEs from heavy 1st/2nd generation of sfermions lead to tachyonic stops unless the latter have a large soft terms at
the UV scale in tension with naturalness~\cite{ArkaniHamed:1997ab}. The size of this effect can be understood by expressing the stop masses in IR
as a function of the UV soft terms;
\begin{align}
 \left(m_{Q_3}^2\right)_{\rm IR}^{\rm MSSM} &\approx -0.11 m_{H_u}^2 + 0.89 m_{Q_3}^2 - 0.08 m_{U_3}^2  - 0.03 m_0(1,2)^2 \nonumber \\ 
 & 3.4 M_3^2 + 0.33
M_2^2
-0.04 M_2 M_3 \,, \\
\label{eq:coeff_mQMSSM}
 \left(m_{U_3}^2\right)_{\rm IR}^{\rm MSSM} &\approx -0.16 m_{H_u}^2 - 0.17 m_{Q_3}^2 + 0.73 m_{U_3}^2  - 0.02 m_0(1,2)^2 \nonumber \\ 
 & 2.9 M_3^2 - 0.12
M_2^2
-0.08 M_2 M_3 
\end{align}
in the MSSM and
\begin{align}
 \left(m_{Q_3}^2\right)_{\rm IR} &\approx -0.07 m_{H_u}^2 + 0.87 m_{Q_3}^2 - 0.03 m_{U_3}^2  - 0.02 m_0(1,2)^2 \nonumber \\ 
 & 3.4 M_3^2 + 11.3
M_X^2
-0.45 M_X M_3 \,, \\
\label{eq:coeff_mQ}
 \left(m_{U_3}^2\right)_{\rm IR} &\approx -0.06 m_{H_u}^2 - 0.05 m_{Q_3}^2 + 0.84 m_{U_3}^2  - 0.02 m_0(1,2)^2 \nonumber \\ 
 & 2.9 M_3^2 - 2.5
M_X^2
-0.93 M_X M_3 
\end{align}
in the model with the new gauge interaction. We see that at one-loop level stops are mainly renormalized by the gluino mass. However, for the
mass of the 1st/2nd generation of sfermions of ${\mathcal{O}}(10)$~TeV the 2-loop effect may dominate. Notice also that this 2-loop effect in the
stops RGEs feeds into the RG running of $m_{H_u}^2$. This is the reason for the positive coefficients in front of  $m_0(1,2)^2$ in
eqs.~\eqref{eq:coeff_mHuMSSM}-\eqref{eq:coeff_mHu} and it may result in destabilization of the EW scale if the 1st/2nd generation of sfermions are too
heavy~\cite{Badziak:2012rf}.

In the right panel of fig.~\ref{fig:ft_M3mstop} we show contours of $\Delta$ for heavy first two generations of sfermions with their masses set to
10~TeV. We see that fine-tuning in the MSSM is always worse than 1~\%. The reason is that even for light stops the soft masses of stops must be large
at the mediation scale to compensate for the negative contribution from the heavy first two generation. This is true to large extent
also  for the
model with the new gauge interaction%
\footnote{Notice also that the coefficient in eq.~\eqref{eq:coeff_mQ} in front of $m_0(1,2)^2$ is smaller than in 
eq.~\eqref{eq:coeff_mQMSSM} for the MSSM which is due to the absence of the two-loop renormalization by the $SU(2)_L$ gauge coupling for $m_{Q_3}^2$.
For a given stop mass this results in smaller UV value of $m_{Q_3}^2$ than in the MSSM, hence smaller tuning. 
} 
but the sensitivity of the EW scale to the soft stop masses is smaller due to the suppressed top Yukawa
coupling.
In consequence, in the model with the new gauge interaction
the region with tuning better than 1\% is only mildly affected by shifting the maximal value of the stop masses from 1.5 to 1.3 TeV. We should
also note that the threshold corrections to the stop masses from the heavy first two generation of squarks, which we neglected in our analysis, are
generically large and positive~\cite{Pierce:1996zz}.  For the 10 TeV squarks the physical stop masses may be about $20\%$ larger than their tree-level
value, as
was recently emphasized in ref.~\cite{Buckley:2016kvr}. After taking this into account, the range of stop masses with the tuning better than 1\% is
essentially unaffected by heavy 1st/2nd generation of sfermions. 

\begin{figure}[t]
\centering
\includegraphics[clip,width=.48\textwidth]{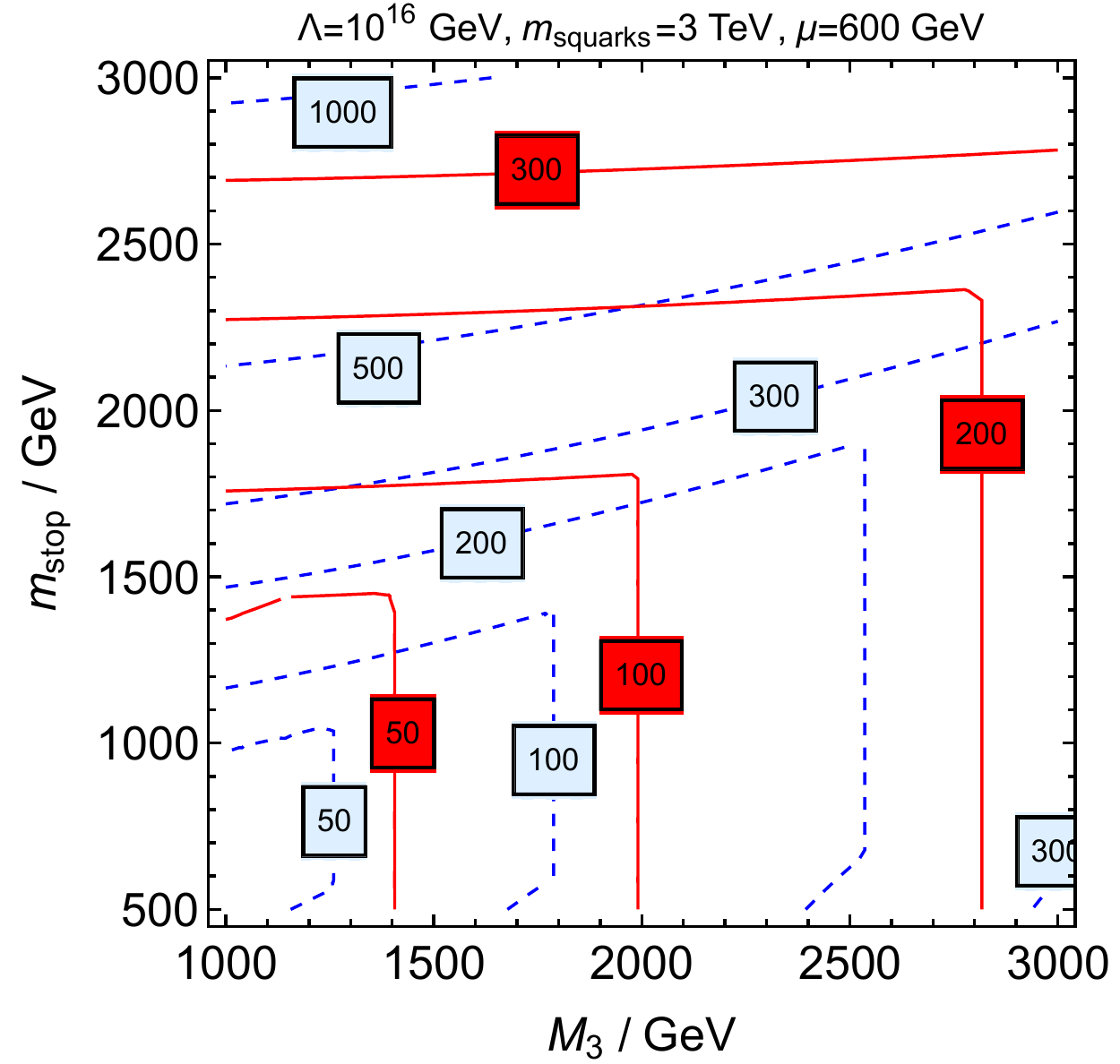}
\includegraphics[clip,width=.48\textwidth]{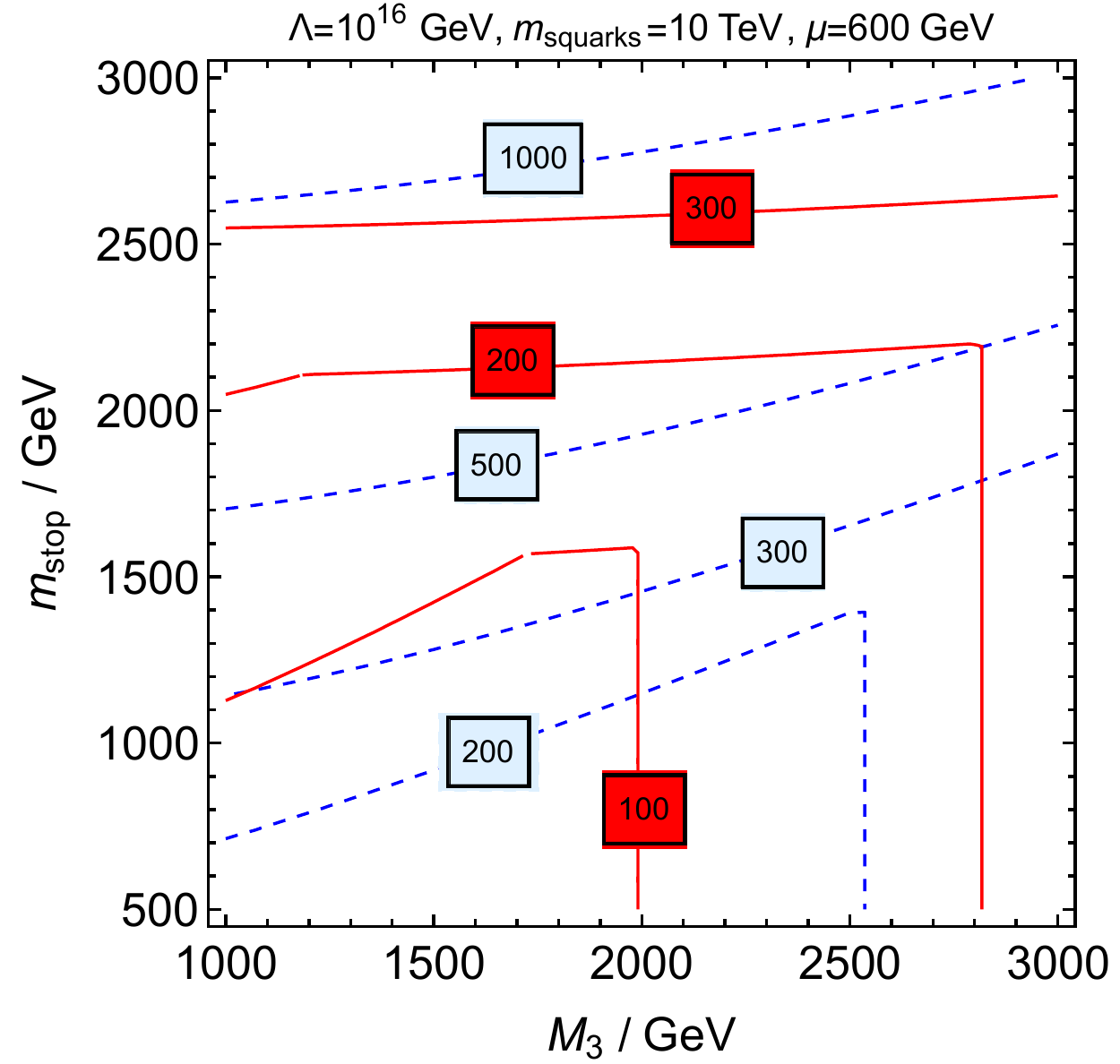}
\caption{
The same as in fig.~\ref{fig:ft_M3mstop} but for $\mu=600$~GeV.
}
\label{fig:ft_M3mstop_mu600}
\end{figure}

The parameters in fig.~\ref{fig:ft_M3mstop} were chosen in such a way that 
$m_{H_u}^2$ is not suppressed at the UV scale so $\Delta_{m_{H_u}^2}$ dominates the total fine-tuning over much of the parameter space. $m_{H_u}^2$ at
the UV scale can
be decreased by increasing $|\mu|$ and/or any of the soft terms on the r.h.s. of eq.~\eqref{eq:coeff_mHu} with positive coefficients. In
fig.~\ref{fig:ft_M3mstop_mu600} contours of $\Delta$ are shown for $\mu=600$~GeV. For such value of $\mu$, $\Delta_{\mu^2}$ is below 50 so does not
affect $\Delta$ (except for gluino mass close to 1 TeV which is excluded anyway for typical SUSY spectra). We see that this choice improves the
fine-tuning by a factor of two in comparison with $\mu=200$~GeV in some of the parameter space and shifts the upper bound on the stop masses with the tuning better than 1\%
up to 1.8 (1.6) TeV for the mass of the first two generation of sfermions set to 3 (10) TeV. 
Similar effect on the fine-tuning is also present in the MSSM so the tuning is still improved
by up to a factor of three.

We have neglected so far the impact of the Higgs mass constraint in the comparison of the fine-tuning of the present model to the MSSM. In our
analysis we assumed vanishing  $A_t$ at the UV scale which generically leads to negligible stop mixing.  In the MSSM with
vanishing stop mixing the 125 GeV Higgs mass implies the stop masses of at least 10~TeV~\cite{susyHD} which leads to $\Delta\gtrsim10^4$. The mass
of stops required to obtain the correct Higgs mass in the MSSM can be reduced to about 2~TeV if the contribution to the Higgs mass from the stop
mixing is maximized which occurs for $|A_t|\approx\sqrt{6}m_{\rm stop}$ at the SUSY scale. The maximal stop mixing is possible even for $A_t=0$ at
the UV scale if $A_t$ is driven to large negative value via strong RG effects from heavy gluino while stops are kept relatively light due to heavy
first two generations of sfermions~\cite{Badziak:2012rf}. This effect can be illustrated by expressing the IR value of $A_t$ in terms of the UV value
of itself and gluino mass in the MSSM:
\begin{equation}
 \left(A_t\right)_{\rm IR}^{\rm MSSM}\approx -1.5 M_3 + 0.3 A_t \,.
\end{equation}
However, in such a scenario 2~TeV stops with $A_t\approx-\sqrt{6}m_{\rm stop}$ require gluino mass of about 7~TeV which leads to
$\Delta\approx1500$. Switching on negative $A_t$ at the UV scale does not help much in reducing $\Delta$ since it has relatively small effect on its
IR value while it gives large correction to $m_{H_u}^2$ via the top Yukawa coupling and easiely dominates tuning if it is large. Thus, in any case
$\Delta$ in the MSSM is at least $1000$ once the Higgs mass constraint is taken into account. We conclude that the model with the extra gauge
interaction improves naturalness by one to two orders of magnitude as compared to the MSSM.

The fine-tuning depends somewhat on explicit implementations of the model that introduces an extra gauge interaction. If
the right-handed top is charged under new gauge group instead of the left-handed one, as it is the case e.g.~in a model proposed in
ref.~\cite{Badziak:2017kjk}, we expect that fine-tuning would be further improved. This is because in such a case the two-loop correction does not
affect $m_{Q_3}^2$ which dominates the tuning from stops in the MSSM and in the present model. The fine-tuning could be also affected by a change in
the number of flavors charged under the extra gauge group and the rank of that gauge group.

\section{Summary and discussion}
\label{sec:concl}

We investigated the impact of an extra gauge interaction on the fine-tuning of the EW scale in supersymmetry. We found that if the new
gauge coupling is large the fine-tuning from stops and gluino is reduced due to suppression of the top Yukawa coupling at higher scales. This effect
is present in any model in which the Higgs and the top quark are charged under the new gauge symmetry and the extra gauge coupling is large. 

We quantified the fine-tuning in an explicit
model with $SU(2)_X \times SU(2)_W$ gauge symmetry which is broken to the SM $SU(2)_L$ gauge group at a scale of around 10~TeV. We found that for the
mediation scale of SUSY breaking of $10^{16}$~GeV the upper bound on the gluino mass from naturalness is increased by few hundred GeV, as
compared to the MSSM, e.g.~for tuning better than 1~\% the maximal value of the gluino mass shifts from 1.8 to 2 TeV. Stops up to about 2 TeV may
also have tuning better than 1~\%.  For stops the improvement is more significant, especially in the scenario with the first two generations
of sfermions much heavier than stops, and the fine-tuning may be a factor of three smaller than in the MSSM with the same stop and gluino masses.
Morevoer, once the Higgs mass constraint is taken into account the fine-tuning in the present model is one to two orders of magnitude smaller than in
the MSSM.

Our findings also demonstrate that it is not possible to find model-independent bounds on stop and gluino masses from naturalness consideration
(even if one sticks to one fine-tuning measure) and in particular, the mechanism that raises the tree-level Higgs mass above that predicted in the MSSM
may have non-negligible impact on the fine-tuning. 

\section*{Acknowledgments}
MB would like to thank the Berkeley Center of Theoretical Physics for its hospitality during completion of this work.
This work has been partially supported by National Science Centre, Poland, under research grants no. 2017/26/D/ST2/00225 and DEC-2014/15/B/ST2/02157 (MB),
by the Office of High Energy Physics of the U.S. Department of Energy under Contract DE-AC02-05CH11231 (KH), and by the National Science Foundation under grant PHY-1316783 (KH).


\begin{thebibliography}{99}


\bibitem{MaianiLecture}
L.~Maiani. in Proceedings: Summer School on Particle Physics, Paris,
	France (1979).
\bibitem{Veltman:1980mj} 
  M.~J.~G.~Veltman,
  Acta Phys.\ Polon.\ B {\bf 12}, 437 (1981).
\bibitem{Witten:1981nf} 
  E.~Witten,
  Nucl.\ Phys.\ B {\bf 188}, 513 (1981).
\bibitem{Kaul:1981wp} 
  R.~K.~Kaul,
  Phys.\ Lett.\ B {\bf 109}, 19 (1982).

\bibitem{Higgsmass_exp}
  G.~Aad {\it et al.} [ATLAS and CMS Collaborations],
  Phys.\ Rev.\ Lett.\  {\bf 114} (2015) 191803
  [arXiv:1503.07589 [hep-ex]].



\bibitem{reviewEllwanger}
  U.~Ellwanger, C.~Hugonie and A.~M.~Teixeira,
  Phys.\ Rept.\  {\bf 496} (2010) 1
  [arXiv:0910.1785 [hep-ph]].

 \bibitem{Ellwanger:2011aa}
  U.~Ellwanger,
  JHEP {\bf 1203} (2012) 044
  [arXiv:1112.3548 [hep-ph]].
  
\bibitem{Hall:2011aa}
  L.~J.~Hall, D.~Pinner and J.~T.~Ruderman,
  JHEP {\bf 1204} (2012) 131
  [arXiv:1112.2703 [hep-ph]].
 
\bibitem{Cao:2012fz}
  J.~J.~Cao, Z.~X.~Heng, J.~M.~Yang, Y.~M.~Zhang and J.~Y.~Zhu,
  JHEP {\bf 1203} (2012) 086
  [arXiv:1202.5821 [hep-ph]].
 

  
\bibitem{Jeong:2012ma}
  K.~S.~Jeong, Y.~Shoji and M.~Yamaguchi,
  JHEP {\bf 1209} (2012) 007
  [arXiv:1205.2486 [hep-ph]].
  
\bibitem{Agashe:2012zq}
  K.~Agashe, Y.~Cui and R.~Franceschini,
  JHEP {\bf 1302} (2013) 031
  [arXiv:1209.2115 [hep-ph]].
  
\bibitem{NaturalNMSSM_King}
  S.~F.~King, M.~Muhlleitner, R.~Nevzorov and K.~Walz,
  Nucl.\ Phys.\ B {\bf 870} (2013) 323
  [arXiv:1211.5074 [hep-ph]].

\bibitem{Gherghetta:2012gb}
  T.~Gherghetta, B.~von Harling, A.~D.~Medina and M.~A.~Schmidt,
  JHEP {\bf 1302} (2013) 032
  [arXiv:1212.5243 [hep-ph]].

\bibitem{Badziak:2013bda}
  M.~Badziak, M.~Olechowski and S.~Pokorski,
  JHEP {\bf 1306} (2013) 043
  [arXiv:1304.5437 [hep-ph]].






  
\bibitem{Batra:2003nj}
  P.~Batra, A.~Delgado, D.~E.~Kaplan and T.~M.~P.~Tait,
  JHEP {\bf 0402} (2004) 043
  [hep-ph/0309149].


\bibitem{Endo:2011gy} 
  M.~Endo, K.~Hamaguchi, S.~Iwamoto, K.~Nakayama and N.~Yokozaki,
  Phys.\ Rev.\ D {\bf 85}, 095006 (2012)
  [arXiv:1112.6412 [hep-ph]].

\bibitem{Cheung:2012zq} 
  C.~Cheung and H.~L.~Roberts,
  JHEP {\bf 1312}, 018 (2013)
  [arXiv:1207.0234 [hep-ph]].

\bibitem{Huo:2012tw} 
  R.~Huo, G.~Lee, A.~M.~Thalapillil and C.~E.~M.~Wagner,
  Phys.\ Rev.\ D {\bf 87}, no. 5, 055011 (2013)
  [arXiv:1212.0560 [hep-ph]].

\bibitem{DAgnolo:2012vzj} 
  R.~T.~D'Agnolo, E.~Kuflik and M.~Zanetti,
  JHEP {\bf 1303}, 043 (2013)
  [arXiv:1212.1165 [hep-ph]].

\bibitem{Craig:2012bs} 
  N.~Craig and A.~Katz,
  JHEP {\bf 1305}, 015 (2013)
  [arXiv:1212.2635 [hep-ph]].

\bibitem{Bertuzzo:2014sma} 
  E.~Bertuzzo and C.~Frugiuele,
  Phys.\ Rev.\ D {\bf 93}, no. 3, 035019 (2016)
  [arXiv:1412.2765 [hep-ph]].

\bibitem{Capdevilla:2015qwa} 
  R.~M.~Capdevilla, A.~Delgado and A.~Martin,
  Phys.\ Rev.\ D {\bf 92}, no. 11, 115020 (2015)
  [arXiv:1509.02472 [hep-ph]].
  
\bibitem{Buckley:2016kvr} 
  M.~R.~Buckley, D.~Feld, S.~Macaluso, A.~Monteux and D.~Shih,
  JHEP {\bf 1708}, 115 (2017)
  [arXiv:1610.08059 [hep-ph]].
\bibitem{Buckley:2016tbs}
  M.~R.~Buckley, A.~Monteux and D.~Shih,
  JHEP {\bf 1706} (2017) 103
  [arXiv:1611.05873 [hep-ph]].


\bibitem{Dimopoulos:2014aua}
  S.~Dimopoulos, K.~Howe and J.~March-Russell,
  Phys.\ Rev.\ Lett.\  {\bf 113} (2014) 111802
  [arXiv:1404.7554 [hep-ph]].
  
 
\bibitem{Garcia:2015sfa}
  I.~Garcia Garcia, K.~Howe and J.~March-Russell,
  JHEP {\bf 1512} (2015) 005
  [arXiv:1510.07045 [hep-ph]].

\bibitem{Fox:2002bu} 
  P.~J.~Fox, A.~E.~Nelson and N.~Weiner,
  JHEP {\bf 0208}, 035 (2002)
  [hep-ph/0206096].

\bibitem{Arvanitaki:2013yja} 
  A.~Arvanitaki, M.~Baryakhtar, X.~Huang, K.~van Tilburg and G.~Villadoro,
  JHEP {\bf 1403}, 022 (2014)
  [arXiv:1309.3568 [hep-ph]].

  
    

\bibitem{Craig:2013fga}
  N.~Craig and K.~Howe,
  JHEP {\bf 1403} (2014) 140
  [arXiv:1312.1341 [hep-ph]].

\bibitem{Katz:2016wtw}
  A.~Katz, A.~Mariotti, S.~Pokorski, D.~Redigolo and R.~Ziegler,
  JHEP {\bf 1701} (2017) 142
  [arXiv:1611.08615 [hep-ph]].

\bibitem{Badziak:2017syq}
  M.~Badziak and K.~Harigaya,
  JHEP {\bf 1706} (2017) 065
  [arXiv:1703.02122 [hep-ph]].

\bibitem{Badziak:2017kjk} 
  M.~Badziak and K.~Harigaya,
  JHEP {\bf 1710}, 109 (2017)
  [arXiv:1707.09071 [hep-ph]].

\bibitem{Badziak:2017wxn}
  M.~Badziak and K.~Harigaya,
  Phys.\ Rev.\ Lett.\  {\bf 120} (2018) 211803
  [arXiv:1711.11040 [hep-ph]].

\bibitem{Chacko:2005pe}
  Z.~Chacko, H.~S.~Goh and R.~Harnik,
  Phys.\ Rev.\ Lett.\  {\bf 96} (2006) 231802
  [hep-ph/0506256].

\bibitem{Falkowski:2006qq}
  A.~Falkowski, S.~Pokorski and M.~Schmaltz,
  Phys.\ Rev.\ D {\bf 74} (2006) 035003
  [hep-ph/0604066].


\bibitem{Chang:2006ra}
  S.~Chang, L.~J.~Hall and N.~Weiner,
  Phys.\ Rev.\ D {\bf 75} (2007) 035009
  [hep-ph/0604076].

\bibitem{susyHD}
  J.~Pardo Vega and G.~Villadoro,
  JHEP {\bf 1507} (2015) 159
  [arXiv:1504.05200 [hep-ph]].
  
\bibitem{Carena:1993bs}
  M.~Carena, M.~Olechowski, S.~Pokorski and C.~E.~M.~Wagner,
  Nucl.\ Phys.\ B {\bf 419} (1994) 213
  [hep-ph/9311222].


\bibitem{Chankowski:1997zh}
  P.~H.~Chankowski, J.~R.~Ellis and S.~Pokorski,
  Phys.\ Lett.\ B {\bf 423} (1998) 327
  [hep-ph/9712234].
 
 
\bibitem{Feng:1999mn}
  J.~L.~Feng, K.~T.~Matchev and T.~Moroi,
  Phys.\ Rev.\ Lett.\  {\bf 84} (2000) 2322
  [hep-ph/9908309].
  
  \bibitem{Abe:2007kf}
  H.~Abe, T.~Kobayashi and Y.~Omura,
  Phys.\ Rev.\ D {\bf 76} (2007) 015002
  [hep-ph/0703044 [HEP-PH]].
  
  \bibitem{Horton:2009ed}
  D.~Horton and G.~G.~Ross,
  Nucl.\ Phys.\ B {\bf 830} (2010) 221
  [arXiv:0908.0857 [hep-ph]].
  
  \bibitem{Ross:2011xv}
  G.~G.~Ross and K.~Schmidt-Hoberg,
  Nucl.\ Phys.\ B {\bf 862} (2012) 710
  [arXiv:1108.1284 [hep-ph]].
  
\bibitem{Yanagida:2013ah} 
  T.~T.~Yanagida and N.~Yokozaki,
  Phys.\ Lett.\ B {\bf 722}, 355 (2013)
  [arXiv:1301.1137 [hep-ph]].
  
  \bibitem{Kaminska:2013mya}
  A.~Kaminska, G.~G.~Ross and K.~Schmidt-Hoberg,
  JHEP {\bf 1311} (2013) 209
  [arXiv:1308.4168 [hep-ph]].

\bibitem{Harigaya:2015iva} 
  K.~Harigaya, T.~T.~Yanagida and N.~Yokozaki,
  PTEP {\bf 2015}, no. 8, 083B03 (2015)
  [arXiv:1504.02266 [hep-ph]].

\bibitem{Harigaya:2015jba} 
  K.~Harigaya, T.~T.~Yanagida and N.~Yokozaki,
  Phys.\ Rev.\ D {\bf 92}, no. 3, 035011 (2015)
  [arXiv:1505.01987 [hep-ph]].

\bibitem{Ross:2017kjc}
  G.~G.~Ross, K.~Schmidt-Hoberg and F.~Staub,
  JHEP {\bf 1703} (2017) 021
  [arXiv:1701.03480 [hep-ph]].

  


\bibitem{Dine:1993np}
  M.~Dine, R.~G.~Leigh and A.~Kagan,
  Phys.\ Rev.\ D {\bf 48} (1993) 4269
  [hep-ph/9304299].
  
\bibitem{Pomarol:1995xc}
  A.~Pomarol and D.~Tommasini,
  Nucl.\ Phys.\ B {\bf 466} (1996) 3
  [hep-ph/9507462].
  
\bibitem{Dudas:1995eq}
  E.~Dudas, S.~Pokorski and C.~A.~Savoy,
  Phys.\ Lett.\ B {\bf 369} (1996) 255
  [hep-ph/9509410].
  
  \bibitem{Barbieri:1995uv}
  R.~Barbieri, G.~R.~Dvali and L.~J.~Hall,
  Phys.\ Lett.\ B {\bf 377} (1996) 76
  [hep-ph/9512388].
 
\bibitem{Dudas:1996fe} 
  E.~Dudas, C.~Grojean, S.~Pokorski and C.~A.~Savoy,
  Nucl.\ Phys.\ B {\bf 481}, 85 (1996)
  [hep-ph/9606383].

 
\bibitem{Cohen:1996vb}
  A.~G.~Cohen, D.~B.~Kaplan and A.~E.~Nelson,
  Phys.\ Lett.\ B {\bf 388} (1996) 588
  [hep-ph/9607394].
  
  

\bibitem{ArkaniHamed:1997ab}
  N.~Arkani-Hamed and H.~Murayama,
  Phys.\ Rev.\ D {\bf 56} (1997) R6733
  [hep-ph/9703259].
  
\bibitem{Badziak:2012rf}
  M.~Badziak, E.~Dudas, M.~Olechowski and S.~Pokorski,
  JHEP {\bf 1207} (2012) 155
  [arXiv:1205.1675 [hep-ph]].
  
\bibitem{Pierce:1996zz}
  D.~M.~Pierce, J.~A.~Bagger, K.~T.~Matchev and R.~j.~Zhang,
  Nucl.\ Phys.\ B {\bf 491} (1997) 3
  [hep-ph/9606211].
  
  
\end{thebibliography}
\end{document}